\documentclass[aps,prd,twocolumn,10pt,superscriptaddress,nofootinbib,nobibnotes,longbibliography]{revtex4-1}

\usepackage{amssymb}
\usepackage{graphicx}
\usepackage{amsmath}
\usepackage{hyperref}
\usepackage{subfigure}
\usepackage{multirow}
\usepackage{setspace}
\usepackage{verbatim}
\usepackage{float}
\usepackage{color}
\usepackage{ulem}
\usepackage[utf8]{inputenc}
\usepackage[table,xcdraw]{xcolor}
\usepackage{makecell}
\usepackage{url}
\usepackage{bm}

\begin{document}
	
%\title{Cosmic Tardigrades: Inflation with Nonminimal Derivative Coupling Surviving Multiple Cosmological Constraints}

%\title{Cosmic tardigrades: A new attractor model near the Harrison-Zeldovich spectrum}

\title{Harrison-Zeldovich attractor: From Planck to ACT results}

\author{Chengjie Fu}
\email[]{fucj@ahnu.edu.cn}
\affiliation{Department of Physics, Anhui Normal University, Wuhu, Anhui 241002, China}

\author{Di Lu}
\affiliation{Department of Physics, Anhui Normal University, Wuhu, Anhui 241002, China}

\author{Shao-Jiang Wang} 
\email[Corresponding author:~]{schwang@itp.ac.cn}
\affiliation{Institute of Theoretical Physics, Chinese Academy of Sciences, Beijing 100190, China }
\affiliation{Asia Pacific Center for Theoretical Physics (APCTP), Pohang 37673, Korea}
	
\begin{abstract}
In the era of Planck cosmology, the inflationary paradigm is best fitted toward the cosmological attractor scenarios, including the induced inflation, universal attractors, conformal attractors, and special attractors that are cataloged as $\xi$-models and $\alpha$-models. The recent hint from the ACT results pushes the scalar spectral index closer to the scale-invariant Harrison-Zeldovich spectrum, calling for a theoretical paradigm shift toward a Harrison-Zeldovich attractor, which is difficult to realize in the standard single-field slow-roll inflationary scenario. In this work, we achieve the Harrison-Zeldovich attractor scenario via nonminimal derivative coupling, attracting the monomial inflation, hilltop inflation, and $\alpha$-attractor E-model toward the Harrison-Zeldovich spectrum.
\end{abstract}

\maketitle

%%%%%%%%%%%%%%%%%%%%%%
\section{Introduction}
%%%%%%%%%%%%%%%%%%%%%%
The inflationary paradigm~\cite{Brout:1977ix,Starobinsky:1980te,Kazanas:1980tx,Sato:1980yn,Guth:1980zm,Linde:1981mu,Albrecht:1982wi,Linde:1983gd} has become the dominant framework for describing the very early Universe and for furnishing natural initial conditions for the hot big bang phase at both the background and perturbation levels. One of the key predictions of the inflationary scenario is the generation of primordial scalar perturbations characterized by a nearly scale-invariant power spectrum~\cite{Mukhanov:1981xt,Mukhanov:1982nu,Hawking:1982cz,Guth:1982ec,Starobinsky:1982ee,Bardeen:1983qw,Kodama:1984ziu,Mukhanov:1985rz}, which is fully consistent with the observations of cosmic microwave background (CMB) by the Planck Collaboration~\cite{Planck:2018vyg,Planck:2018jri}. The Planck results determine the scalar spectral index $n_s$ to be $n_s=0.9649\pm0.0042$~\cite{Planck:2018jri}.
Another crucial prediction of inflation is the production of primordial tensor perturbations (primordial gravitational waves)~\cite{Sahni:1990tx,Allen:1987bk}, which remain undetected and are being sought via measurements of B-mode polarization in the CMB. A joint analysis of Planck and BICEP/Keck data yields a stringent upper limit on the tensor-to-scalar ratio $r$ with $r<0.036$~\cite{BICEP:2021xfz}. 

The combined constraints on $n_s$ and $r$ can be used to rule out competing inflationary models, such as the monomial inflation~\cite{Linde:1983gd,Silverstein:2008sg,McAllister:2014mpa} and natural inflation~\cite{Freese:1990rb,Adams:1992bn}. The best-fit possibility comes from a single-field slow-roll inflation~\cite{Linde:1983gd} with a plateau potential of an exponentially flat form at large field values~\cite{Cai:2021png}, which could be described at ultraviolet by some cosmological attractor models~\cite{Galante:2014ifa}, such as the induced inflation\cite{Giudice:2014toa,Pallis:2013yda,Pallis:2014dma,Pallis:2014boa,Kallosh:2014rha}, universal attractors (including Higgs inflation~\cite{Bezrukov:2007ep,Bezrukov:2008ej,Bezrukov:2009db}), conformal attractors (including Starobinsky model~\cite{Starobinsky:1980te,Vilenkin:1985md} and T-model~\cite{Kallosh:2013hoa,Cai:2014bda}), and special attractors (with noncanonical kinetic term) that are classified into $\xi$-models~\cite{Kallosh:2013tua} and $\alpha$-models~\cite{Kallosh:2013hoa,Kallosh:2013daa,Ferrara:2013rsa,Kallosh:2013yoa,Linde:2015uga}. 

Notably, recent CMB observations from the Atacama Cosmology Telescope (ACT) suggest a higher value of $n_s$~\cite{AtacamaCosmologyTelescope:2025blo,AtacamaCosmologyTelescope:2025nti}. Specifically, the combination of Planck, ACT, and Dark Energy Spectroscopic Instrument (DESI) data yields a spectral index of $n_s=0.9743 \pm 0.0034$~\cite{AtacamaCosmologyTelescope:2025blo}. The new results can accommodate the monomial inflation model, but disfavor several previously favored inflationary models, including the Starobinsky $R^2$ inflation model~\cite{AtacamaCosmologyTelescope:2025nti}. Numerous studies have attempted to reconcile a wide range of inflationary models with the ACT data by invoking modified theories of gravity~\cite{Kallosh:2025gmg,Dioguardi:2025vci,Salvio:2025izr,Dioguardi:2025mpp,Gao:2025onc,He:2025bli,Gialamas:2025ofz,Yogesh:2025wak,Addazi:2025qra,Pallis:2025nrv,Odintsov:2025wai,Choudhury:2025vso,Gao:2025viy,Zahoor:2025nuq,Ketov:2025cqg,Zhu:2025twm,Odintsov:2025bmp,Zhang:2025lfx,Odintsov:2025jky,Oikonomou:2025htz,Modak:2025bjv,Choi:2025qot,Odintsov:2025eiv,Qiu:2025uot}, alternative reheating history~\cite{Drees:2025ngb,Zharov:2025evb,Haque:2025uri,Liu:2025qca,Haque:2025uis,Zharov:2025evb,Maity:2025czp,Zharov:2025zjg,Mohammadi:2025gbu,Chen:2025qxq}, and other extensions~\cite{Aoki:2025wld,Gialamas:2025kef,Byrnes:2025kit,Heidarian:2025drk,Wolf:2025ecy,Han:2025cwk,Ahmed:2025sfm,Yuennan:2025kde,Oikonomou:2025xms,Aoki:2025ywt,Yuennan:2025tyx,Pallis:2025vxo,Yuennan:2025inm,Ellis:2025bzi,Allegrini:2025jha,DOnofrio:2025bol,Afshar:2025ndm}. In particular, the inflation models with polynomially flat potentials~\cite{Kallosh:2025ijd} have been shown to provide a good fit to the ACT data in certain regions of parameter space (see, e.g., Refs.~\cite{Yi:2025dms,Peng:2025bws}).

On the other hand, the early dark energy (EDE) models~\cite{Poulin:2018cxd,Lin:2019qug,Smith:2019ihp,Niedermann:2019olb,Sakstein:2019fmf,Ye:2020btb,Gogoi:2020qif,Braglia:2020bym,Ye:2021iwa,Karwal:2021vpk,Wang:2022jpo,Poulin:2023lkg} have emerged as a promising framework for alleviating the Hubble tension by shrinking the sound horizon~\cite{Ye:2021nej}. After including recent ACT data, EDE remains a potential resolution to the Hubble tension~\cite{Poulin:2025nfb}. Introducing EDE typically shifts $n_s$ toward unity, implying a Harrison–Zeldovich spectrum with $n_s=1$~\cite{Ye:2020btb,Jiang:2022uyg,Jiang:2022qlj,Wang:2024tjd}. This result has been reconfirmed by fitting the anti de Sitter (AdS)-EDE model~\cite{Ye:2020btb} to ACT data combined with Planck and South Pole Telescope (SPT), yielding $n_s= 0.9960 \pm 0.0047$~\cite{Peng:2025tqt}. Realizing such an extremely scale-invariant spectrum within the single-field slow-roll inflationary paradigm would call for an unrealistic {\it e}-folding number, which can be surprisingly avoided for a monomial potential via a negative nonminimal derivative coupling (NDC)~\cite{Fu:2023tfo}. See also Refs.~\cite{Takahashi:2021bti,Ye:2022efx,Braglia:2022phb} for multifield realizations of the Harrison–Zeldovich spectrum from axion curvaton and hybrid waterfall models, respectively.

Therefore, the same trend in favor of a more scale-invariant spectrum shows up in both ACT results and EDE scenarios, but significant uncertainties persist in the observational constraints on $n_s$ and $r$. It is therefore desirable to develop a theoretical framework that can render a broad class of inflationary models flexibly compatible with diverse observational constraints. In this work, we construct such a framework by employing the mechanism of gravitationally enhanced or weakened friction arising from an NDC, leading to a new cosmological attractor toward the Harrison–Zeldovich spectrum.

%%%%%%%%%%%%%%%%%%%%%%
\section{Inflation with nonminimal derivative coupling}
%%%%%%%%%%%%%%%%%%%%%%
The NDC model is described by the Lagrangian:
\begin{align}\label{action}
 \mathcal{L}=\frac{M_{\rm Pl}^2}{2}R -\frac{1}{2}\left(g_{\mu\nu}- \xi G_{\mu\nu}\right)\nabla^\mu\phi\nabla^\nu\phi- V(\phi),
\end{align}
where $M_{\rm Pl}$ denotes the reduced Planck mass, $g_{\mu\nu}$ is the metric tensor, and $G_{\mu\nu}$ is the Einstein tensor. The parameter $\xi$ has dimension of mass$^{-2}$. This model is a member of Horndeski's theory (also known as generalized Galileons), which represents the most general scalar-tensor theory with second-order field equations. The nonstandard kinetic term $G^{\mu\nu}\nabla_\mu\phi\nabla_\nu\phi$ emerges from the term $G_5(\phi,X)G_{\mu\nu}\nabla^\mu\nabla^\nu\phi$ with $X=-\partial_\mu\phi\partial^\mu\phi/2$ in the generalized Galileons, after performing integration by parts. Consequently, one finds $\xi\propto{\rm d}G_5/{\rm d}\phi$, and the coupling parameter $\xi=\xi(\phi,X)$ can be regarded as an arbitrary function of $\phi$ and $X$. In this paper, we focus on the case where $\xi$ is treated as a function of $\phi$ alone.

Considering that the scalar field $\phi$ acts as the inflaton driving cosmic inflation, we work within the spatially flat FRW background, characterized by the line element ${\rm d}s^2=-{\rm d}t^2 + a(t)^2\delta_{ij}{\rm d}x^i{\rm d}x^j$ with $a(t)$ being the scale factor. The inflationary dynamics in this setup are governed by the following equations:
\begin{align}
\label{BG1} 3M_{\rm Pl}^2 H^2 = &\frac{1}{2}\left( 1 + 9\xi H^2\right)\dot\phi^2 + V(\phi), \\[6pt]
\left(1+3\xi H^2\right)\ddot\phi + &\bigg[ 1 + \xi \left( 2\dot H + 3H^2\right)\bigg]3H\dot\phi \nonumber \\
\label{BG2} & \qquad\quad + \frac{3}{2}\xi_{\phi}H^2\dot\phi^2 + V_{\phi} = 0,
\end{align}
where $H$ is the Hubble parameter and a subscript $\phi$ means a derivative with respect to the inflaton field. The evolution of various physical quantities during inflation is typically described with respect to the {\it e}-folding number $N$, defined by $N=\ln(a_{\rm e}/a)$ with $a_{\rm e}$ being the scale factor at the end of inflation. 

After introducing the four slow-roll parameters, $\epsilon=-\dot H/H^2$, $\eta = \ddot\phi/(H\dot\phi)$, $\delta_X = \dot\phi^2/(2M_{\rm Pl}^2 H^2)$, and $\delta_D=\xi\dot\phi^2/(4M_{\rm Pl}^2)$, the background equations can be approximated during inflation as
\begin{align}\label{reduced_BG}
   3M_{\rm Pl}^2 H^2 \simeq  V(\phi), \qquad (1 + 3\xi H^2)3H \dot\phi  + V_{\phi} \simeq 0,
\end{align}
by using the slow-roll conditions $\{\epsilon,|\eta|, \delta_X, |\delta_D|\}\ll 1$ and assuming that the contribution of the $\xi_{\phi}$ term is negligible. Unlike the minimally coupled case, $\epsilon$ and $\delta_X$ are no longer identical but instead satisfy a modified relation, $\epsilon\simeq \delta_X+6\delta_D$, under the slow-roll approximation. In general, neglecting the $\xi_{\phi}$ term requires that the time variation of the coupling parameter be sufficiently slow during inflation, namely $|\dot\xi/(H\xi)|=|\xi_\phi \dot\phi/(H\xi)|\ll1$. This condition is analogous to the slow-roll requirement that $|\eta|\ll1$. It is straightforward to see that the NDC modifies the effective friction in the equation of motion of the inflaton relative to the conventional single-field slow-roll inflation with a minimally coupled and canonical inflaton. In particular, compared with the $\xi=0$ case, the friction is enhanced for $\xi>0$ and diminished for $\xi<0$. It should be emphasized that, in the $\xi<0$ case, the condition $\left(1+3\xi H^2\right)>0$ must be satisfied to avoid instabilities in the system, thereby imposing a lower bound on $\xi$. Given the Planck upper limit on the inflationary Hubble parameter, $H_{\rm inf}<2.5\times10^{-5}M_{\rm Pl}$~\cite{Planck:2018jri}, this condition implies $\xi\gtrsim-5.3\times10^8M_{\rm Pl}^{-2}$.

Since the coupling parameter $\xi$ may evolve with the inflaton, $\xi=\xi(\phi)$, we assume that the NDC is negligible during the early stage of inflation and becomes relevant only at later times. Consequently, the early-time dynamics of $\phi$ coincide with those of conventional single-field slow-roll inflation, whereas at late times the coupling modifies the effective friction, slowing the roll for $\xi>0$ and speeding it up for $\xi<0$. 

As shown in Fig.~\ref{fig1}, the NDC becomes significant when $\phi < \phi_{\rm c}$, where $\phi_{\rm c}$ denotes the critical field value. For $\phi > \phi_{\rm c}$, $\xi \simeq 0$, and the effect of $\xi$ can be neglected. However, when $\phi < \phi_{\rm c}$, the NDC introduces either high or low friction compared to the conventional slow-roll evolution. The field value $\phi_{\rm e}$ at the end of inflation differs from the conventional case: specifically, $\phi_{\rm e}$ decreases for $\xi > 0$ due to the enhanced friction, and increases for $\xi < 0$ due to the reduced friction.

For the $\xi < 0$ case, the inflaton excursion $\Delta \phi = |\phi_{\rm c}- \phi_{\rm e}|$ becomes smaller. Meanwhile, due to the lower friction, the inflaton spends less time traversing this region. As a result, the {\it e}-folding number $N^{(-)}_{\rm c}=\ln(a_{\rm e}/a_{\rm c})$, with the subscript c denoting the evaluation at $\phi = \phi_{\rm c}$, is smaller than the {\it e}-folding number $N_{\rm c}$ of conventional slow-roll inflation, i.e., $N^{(-)}_{\rm c} <  N_{\rm c}$. 

In contrast, for the $\xi > 0$ case, the {\it e}-folding number $ N^{(+)}_{\rm c}$ increases relative to $N_{\rm c}$ due to the larger inflaton excursion $\Delta \phi$ and the enhanced friction. Fixing the {\it e}-folding number $ N_\ast=\ln(a_{\rm e}/a_{\ast})=60$, which corresponds to $\phi = \phi_{\ast}$ with $\phi_{\ast}$ being the field value at the moment when the CMB pivot scale exits the horizon, the field value $\phi_{\ast}$ will change. This is because $\Delta N^{(\pm)}_{\phi_{\ast} \rightarrow \phi_{\rm c}}= N_\ast - N^{(\pm)}_{\rm c} \neq \Delta N_{\phi_{\ast} \rightarrow \phi_{\rm c}}$, and during the period from $\phi=\phi_\ast$ to $\phi=\phi_{\rm c}$, the inflaton experiences the conventional slow-roll evolution. 

For $\xi < 0$, $\phi_\ast$ become larger since $\Delta N^{(-)}_{\phi_{\ast} \rightarrow \phi_{\rm c}} > \Delta N_{\phi_{\ast} \rightarrow \phi_{\rm c}}$, whereas for $\xi>0$, $\phi_\ast$ become smaller since $\Delta N^{(+)}_{\phi_{\ast} \rightarrow \phi_{\rm c}} < \Delta N_{\phi_{\ast} \rightarrow \phi_{\rm c}}$. The foregoing discussion assumed inflationary scenarios in which the inflaton begins at large field values and rolls toward smaller values (e.g., monomial inflation and $\alpha$-attractor E-model). It also applies to models in which the inflaton starts at small field values and evolves toward larger values (e.g., hilltop and natural inflation); in those cases, only the direction of the variation of $\phi_\ast$ is reversed. In all cases, a common feature is that for $\xi < 0$ the value $\phi_{\ast}$ lies farther from the potential minimum, 
whereas for $\xi > 0$ the value $\phi_{\ast}$ lies closer to the potential minimum.

\begin{figure}[htbp]
\centering
\includegraphics[width=1\columnwidth]{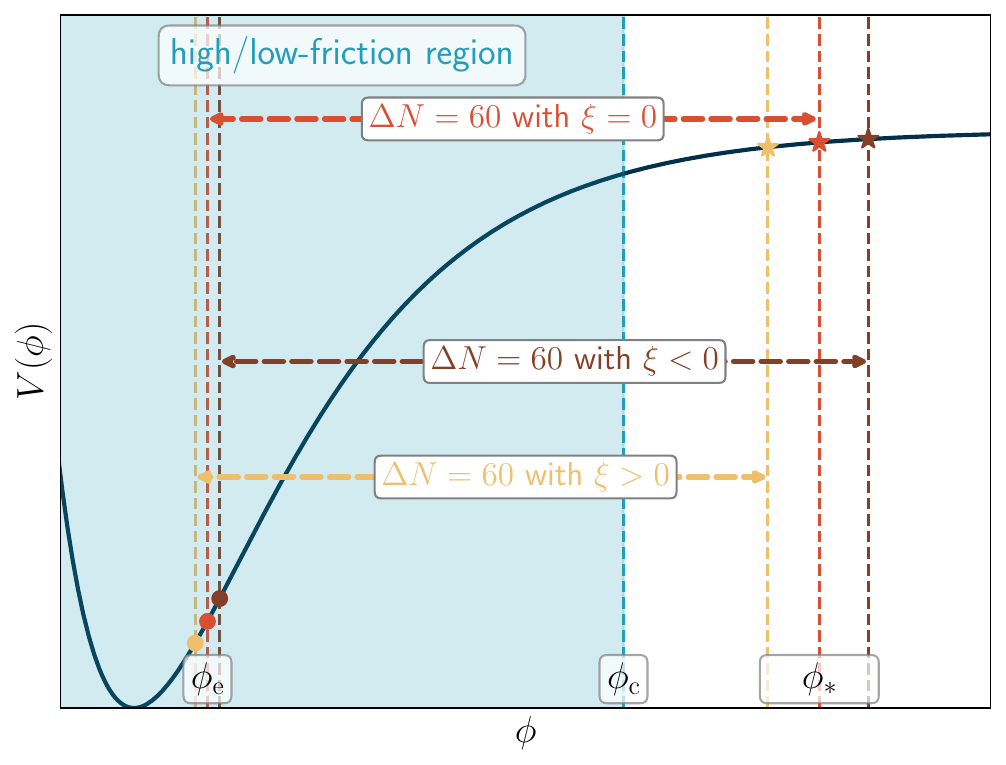}
\caption{\label{fig1}  Schematic illustration of the effect of a high or low friction region on the field value $\phi_\ast$. }
\end{figure}

Since the NDC has little effect around $\phi=\phi_\ast$, the scalar spectral index $n_s$ and tensor-to-scalar ratio $r$ at the CMB scales follow the results of the conventional single-field slow-roll inflation, which are given by
\begin{align}\label{expression_ns_r}
n_s = 1-3 M^2_{\rm Pl}\left( \frac{V_\phi}{V}\right)^2 + 2M^2_{\rm Pl}\frac{V_{\phi\phi}}{V},\;\; r=8 M^2_{\rm Pl}\left( \frac{V_\phi}{V}\right)^2.
\end{align}
In general, $n_s$ and $r$ evaluated at a field value further from the potential minimum tend to be larger and smaller, respectively, since the potential around this field value is flatter. In contrast, $n_s$ and $r$ evaluated at a field value closer to the potential minimum are smaller and larger, respectively, since the potential around this field value is steeper. If $N_\ast=60$ is fixed, the value of $\phi_\ast$ can still vary, as discussed in the above scenario. Consequently, the predicted values of $n_s$ and $r$ at the CMB pivot scale may shift and thus become consistent with the constraints inferred from different datasets. In the next section, we apply this scenario to reconcile several inflationary models with a range of current observational constraints.

%%%%%%%%%%%%%%%%%%%%%%
\section{Observational constraints}
%%%%%%%%%%%%%%%%%%%%%%
To implement the scenario described above, we assume that $\xi(\phi)$ takes the following approximate form,
\begin{align}\label{xi}
    \xi(\phi) = \lambda \Theta[s(\phi-\phi_c)],
\end{align}
where $\lambda$ is a constant with mass$^{-2}$ dimension, and $\Theta$ is the Heaviside step function. Here $s=\pm1$ denotes the rolling direction of the inflaton: $s=-1$ corresponds to evolution from large to small field values, while $s=1$ corresponds to evolution from small to large. For this functional form, $\xi(\phi)$ is constant throughout most of inflation and varies appreciably only within an extremely short interval (around $\phi=\phi_c$). Consequently, the integrated contribution of the $\xi_{\phi}$ term to the background evolution is negligible, and it can be safely omitted in deriving Eq. \eqref{reduced_BG}, as also verified by the agreements with analytical results derived in the Supplemental Material~\cite{SMHZ}.
%For this functional form, the condition that the $\xi_{\phi}$ term is negligible in Eq.~\eqref{BG2} holds throughout most of inflation.

\begin{figure}[htbp]
\centering
\includegraphics[width=1\columnwidth]{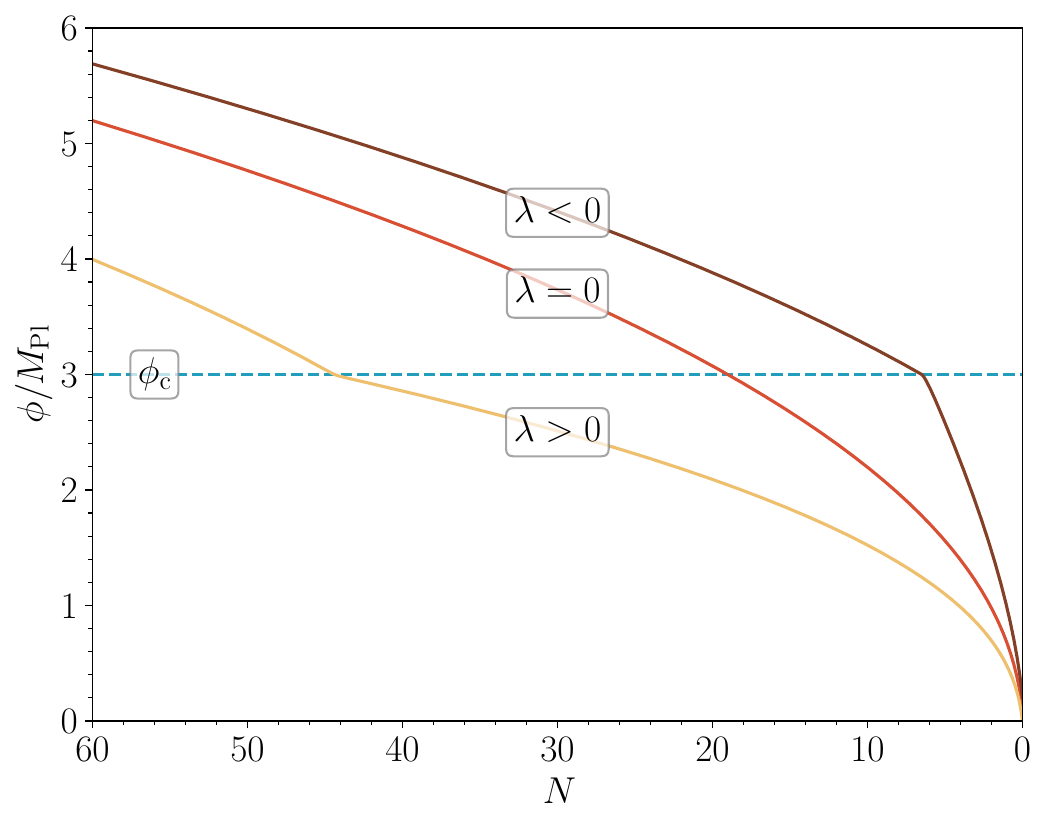}
\caption{\label{fig2}  The evolution of $\phi$ as a function of $N$ for the monomial potential with $p=1/10$, {\bf with $\xi$ given by Eq. \eqref{xi}. The results are shown for $\lambda=0$, $\lambda<0$, and $\lambda>0$}.}
\end{figure}

\begin{figure*}[htbp]
    \centering
    \includegraphics[width=1\textwidth]{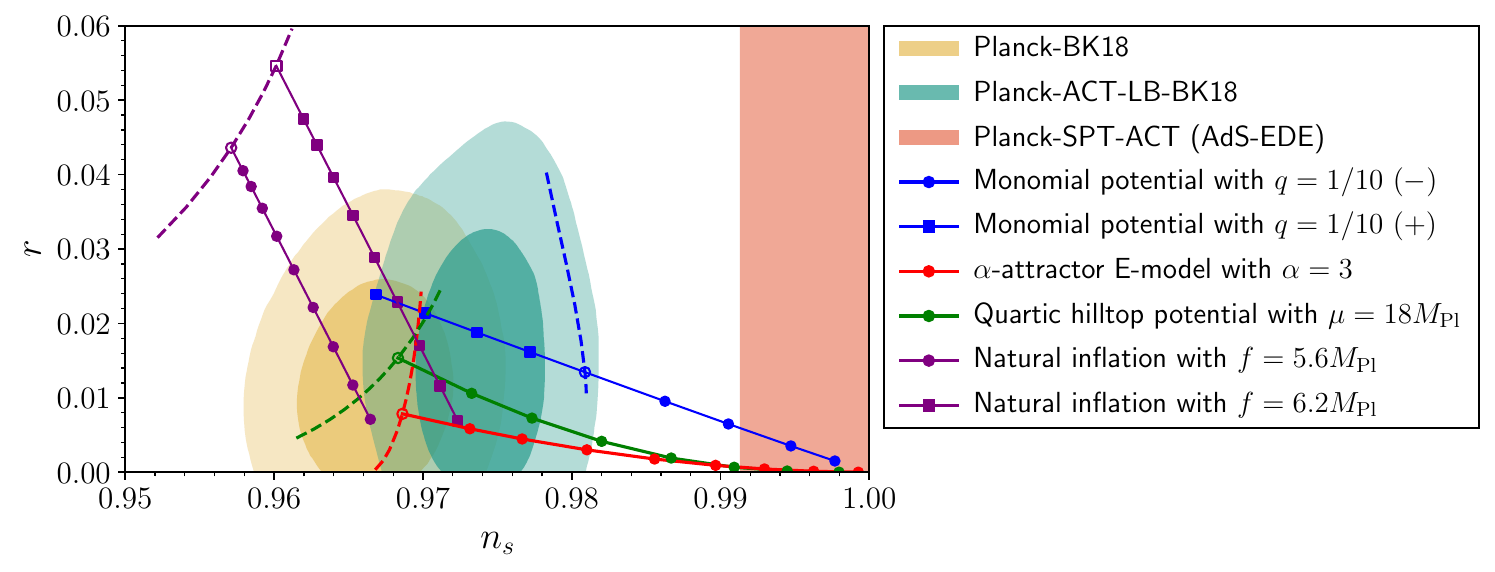}
    \caption{The theoretical predictions in the $n_s-r$ plane for typical inflation models with the help of NDC. The yellow-shaded regions show the constraints from Planck and BICEP/Keck (Planck–BK18). The green-shaded regions denote the joint analysis of Planck and ACT, including CMB lensing and BAO measurements from DESI, together with BICEP/Keck (Planck–ACT–LB–BK18). The red-shaded region indicates the credible interval for $n_s$ inferred from the Planck-SPT-ACT dataset within the AdS-EDE extension of $\Lambda$CDM. The dashed lines and open markers indicate predictions for $n_s$ and $r$ in the minimal coupling case for monomial potential (blue), $\alpha$-attractor E-model (red), quartic hilltop potential (green), and natural inflation (purple). As an illustrative example for the monomial potential, we adopt $q=1/10$ and $M=0.01 M_{\rm Pl}$. For one parameter set we choose $\lambda\Lambda^4M_{\rm Pl}^{-2}=-0.94$ with $\phi_{\rm c}/M_{\rm Pl}=\{4, 6,9,14 \}$ (blue dots); for the other we choose $\lambda\Lambda^4M_{\rm Pl}^{-2}=2$ with $\phi_{\rm c}/M_{\rm Pl}=\{2, 2.5, 2.8, 3 \}$ (blue squares). For the $\alpha$-attractor E-model we adopt $\alpha=3$ and $\lambda\Lambda^4M_{\rm Pl}^{-2} = -1$ with $\phi_{\rm c}/M_{\rm Pl}=\{5, 6, 7, 8, 9, 10, 11.5, 15 \}$ (red dots). For the quartic hilltop potential we adopt $\mu=18M_{\rm Pl}$ and $\lambda\Lambda^4M_{\rm Pl}^{-2} = -1$ with $\phi_{\rm c}/M_{\rm Pl}=\{ 12.5, 11, 9.5, 8, 6.5, 5, 3\}$ (green dots). For the natural inflation we adopt $f=5.6M_{\rm Pl}$ (purple dots) and $f=6.2 M_{\rm Pl}$ (purple squares) with $\lambda\Lambda^4M_{\rm Pl}^{-2} = -0.5$ and $\phi_{\rm c}/M_{\rm Pl}=\{ 11, 10, 9, 8, 7, 6, 5, 4, 3\}$.
    The monomial potential, $\alpha$-attractor E-model, and quartic hilltop potential can be attracted toward the Harrison-Zeldovich spectrum, while the natural inflation seems unlikely. \label{fig3} }
\end{figure*}

%Monomial potentials of the form $V\sim\phi^p$ are already excluded by the Planck and BICEP/Keck observations~\cite{BICEP:2021xfz}. However, the inclusion of ACT data can accommodate the monomial potentials with $p\lesssim2/3$~\cite{ACT:2025tim}. We consider a well-defined version of the monomial potentials, given by
We first consider monomial inflation with the potential $V\sim\phi^p$, and specifically its well-defined form~\cite{Silverstein:2008sg,McAllister:2014mpa},
\begin{align}\label{monomial_potentials}
    V(\phi) = \Lambda^4 \left[ \left(1 + \frac{\phi^2}{M^2}\right)^{p/2} -1\right],
\end{align}
which is bounded from below and has a minimum. Taking the above potential as an example, let us illustrate the inflationary dynamics in the presence of the NDC with the coupling parameter~\eqref{xi}. Monomial inflation corresponds to $s=-1$, and we adopt the parameter set: $p=1/10$, $M=0.01 M_{\rm Pl}$, $\phi_c= 3 M_{\rm Pl}$, and $\lambda\Lambda^4M_{\rm Pl}^{-2}=\{-1,0,2\}$, showing the evolution of $\phi$ as a function of $N$ in Fig.~\ref{fig2}. It is clear that the inflaton excursion from $\phi_{\rm c}$ to $\phi_{\rm e}$ corresponds to fewer {\it e}-folds for the $\lambda < 0$ case and more {\it e}-folds for the $\lambda > 0$ case, compared to the $\lambda = 0$ case. Consequently, the field value $\phi_\ast$, corresponding to the time when $N = 60$, increases for the $\lambda < 0$ case and decreases for the $\lambda > 0$ case. By selecting appropriate values for $\phi_{\rm c}$ and $\lambda$, we can adjust $n_s$ and $r$ at the CMB pivot scale to align with the observationally favored region. As seen from Fig.~\ref{fig3}, the monomial inflation with the NDC can be consistent with the constraints from Planck-BK18, Planck-ACT-LB-BK18, and Planck-SPT-ACT with EDE for the different parameter choices.

The inflationary predictions from the $\alpha$-attractor E-model~\cite{Kallosh:2013hoa},
\begin{align}
    V(\phi) = \Lambda^4 \left( 1 -  e^{-\sqrt{\frac{2}{3\alpha}}\frac{\phi}{M_{\rm Pl}}}\right)^2,
\end{align}
and the quartic hilltop potential~\cite{Boubekeur:2005zm},
\begin{align}
    V(\phi) = \Lambda^4 \left( 1 - \frac{\phi^4}{\mu^4} \right),
\end{align}
are in good agreement with the Planck and BICEP/Keck observations~\cite{BICEP:2021xfz}. The increase in the scalar spectral index resulting from the ACT data leads to significantly reduced parameter ranges for these models, compared to those obtained from the Planck data. Within the NDC framework, the $\alpha$-attractor E-model corresponds to $s=-1$, while the quartic hilltop potential corresponds to $s=1$. The predictions can be better compatible with the constraints from Planck-ACT-LB-BK18, even from Planck-SPT-ACT with EDE, by increasing $n_s$ and decreasing $r$ in the case of $\lambda<0$.

Finally, we focus on the natural inflation with the potential~\cite{Freese:1990rb,Adams:1992bn},
\begin{align}
    V(\phi) = \Lambda^4 \left[ 1 + \cos\left(\frac{\phi}{f}\right) \right],
\end{align}
which is strongly disfavored by the current observational constraints due to the smaller $n_s$. In conventional slow-roll inflation, the scalar spectral index, given in Eq.~\eqref{expression_ns_r}, can be approximated as $n_s\simeq 1 - M_{\rm Pl}^2/f^2$ in the limit of $\phi\ll f$. Unlike the potentials discussed above, even when $\phi_\ast$ lies in the limiting regime $\phi_\ast \ll f$, the resulting $n_s$ remains dependent on the potential parameter $f$. Consequently, even with the NDC for $s=1$ and negative $\lambda$, the compatibility of the model depends sensitively on the choice of $f$.
In particular, $f=5.6M_{\rm Pl}$ is compatible with the results from Planck-BK18, whereas $f=6.2M_{\rm Pl}$ is compatible with both Planck–BK18 and Planck–ACT–LB–BK18. To reproduce $n_s\simeq 1$ as favored by Planck–SPT–ACT with EDE, one requires $f\gtrsim 10 M_{\rm Pl}$. In that case, the inflaton excursion from $\phi_{\ast}$ to $\phi_{\rm e}$ becomes excessively large. 
Given the lower bound on $\xi$ and retaining $N_\ast \sim 60$, $\phi_\ast$ cannot be shifted into the regime $\phi \ll f$. Consequently, the resulting $n_s$ cannot be brought into agreement with the constraints from Planck-SPT-ACT with EDE.

As a consistency check, we analytically derive the expressions for $n_s$ and $r$ in the Supplemental Material~\cite{SMHZ}, and show that the analytical results are in good agreement with the numerical ones. This agreement also supports neglecting the term $\xi_\phi$ in the derivation of Eq.~\eqref{reduced_BG}.

%%%%%%%%%%%%%%%%%%%%%%
\section{Conclusion and discussion}
%%%%%%%%%%%%%%%%%%%%%%

In this work, we have proposed a unified framework to reconcile a broad class of inflationary models with the diverse and sometimes conflicting observation constraints on the scalar spectral index $n_s$ and the tensor-to-scalar ratio $r$. By incorporating an NDC between the inflaton and the Einstein tensor, we introduce a mechanism that effectively modulates the friction experienced by the inflaton during its evolution. Depending on the sign of the coupling parameter $\xi$, the friction can be either enhanced $(\xi>0)$ or weakened $(\xi<0)$, thereby altering the inflationary trajectory and the corresponding field value $\phi_\ast$ at which CMB scales exit the horizon.

We have demonstrated that this mechanism can significantly shift the predictions for $n_s$ and $r$ without altering the underlying potential $V(\phi)$. For $\xi<0$, the reduced friction leads to a larger $\phi_*$, typically yielding a higher $n_s$ and a lower $r$, which better aligns with recent ACT and EDE-favored constraints. Conversely, for $\xi>0$, the enhanced friction results in a smaller $\phi_*$, generally lowering $n_s$ and raising $r$, consistent with more conventional Planck-based bounds.

Through explicit examples, including monomial inflation, $\alpha$-attractor E-model, quartic hilltop inflation, and natural inflation, we have shown that the NDC framework can flexibly accommodate a wide range of observational datasets, such as Planck-BK18, Planck-ACT-LB-BK18, and even Planck-SPT-ACT with EDE. However, we also identified limitations that, in the case of natural inflation, the requirement of a very large decay constant $f\gtrsim10\,M_\mathrm{Pl}$ to achieve $n_s\approx1$ leads to an excessively large field excursion, which cannot be realistically accommodated within the NDC framework while maintaining $N_*\sim60$.

Our results highlight the potential of gravitational friction modulation as a powerful and generic tool for adapting inflationary models to evolving observational data, including the approach to a new cosmological attractor characterized by a scale-invariant power spectrum. Future work could explore more general forms of $\xi(\phi,X)$, incorporate reheating dynamics, or extend the analysis to multifield scenarios. Moreover, as next-generation CMB experiments of $n_s$ and $r$, the flexibility offered by the NDC mechanism may prove essential in bridging the gap between theory and observation.

We note that the present model may be degenerate with other effects operating during or after inflation. During inflation, additional friction sourced by the backreaction from particle production can lead to a qualitatively similar effect \cite{Bastero-Gil:2021fac,Bastero-Gil:2026ypn}. In addition, a nonstandard postinflationary history can also be degenerate with our mechanism~\cite{Drees:2025ngb,Zharov:2025zjg,Haque:2025uri,Liu:2025qca,Maity:2025czp,Mondal:2025kur,Chakraborty:2025oyj,Chen:2025qxq,Ellis:2025bzi,Ellis:2025zrf}, since the value of the {\it e}-folding number $N_\ast$ depends on the reheating history~\cite{Dai:2014jja,Creminelli:2014fca,Martin:2014nya,Munoz:2014eqa,Cai:2015soa,Cook:2015vqa}. In particular, a reheating phase with equation of state $w<1/3$ decreases $N_\ast$, whereas a phase with  $w>1/3$ increases it~\cite{Creminelli:2014fca}. Therefore, postinflationary reheating can mimic the effects of $\xi>0$ or $\xi<0$ in this model. However, this degeneracy is limited to the Planck- and ACT-favored regions of $n_s$. As shown in the Supplemental Material \cite{SMHZ}, postinflationary reheating can increase $N_\ast$ by at most $\sim14$~\cite{Munoz:2014eqa}, which is insufficient to shift $n_s$ to values close to unity. On the other hand, this can be achieved in the present mechanism. 
 
%%%%%%%%%%%%%%%%%%%%%%%%%%%%%%%%%%%%%%%%%%%%%%%%%%%%%%%%%%%%%%%%%%%%%%%%%%%
\begin{acknowledgments}
We thank Zu-Cheng Chen and Shoulong Li for fruitful discussions. 
This work is supported by the National Natural Science Foundation of China Grants No. 12305057, No. 12422502, No. 12547110, No. 12105344, No. 12588101, No. 12235019, and No. 12447101, 
the National Key Research and Development Program of China Grants No. 2021YFC2203004 and No. 2021YFA0718304, 
and the China Manned Space Program Grant No. CMS-CSST-2025-A01.
\end{acknowledgments}

\textbf{Data availability}: The data are not publicly available. The data are available from the authors upon reasonable request.

%%%%%%%%%%%%%%%%%%%%%%%%%%%
%\bibliographystyle{apsrev4-2}
\bibliography{references}  
%%%%%%%%%%%%%%%%%%%%%%%%%%%

\newpage
\onecolumngrid

\appendix

\section{Supplemental Material}

In this Supplemental Material, we analytically compute the scalar spectral index $n_s$ and the tensor-to-scalar ratio $r$. The {\it e}-folding number from the time $t_\ast$ at which $\phi=\phi_\ast$ to the time $t_{\rm e}$ at the end of inflation is defined by $N_\ast=\int^{t_{\rm e}}_{t_\ast}H(t){\rm d}t$. Using Eqs. (4) and (6) in the main text, we obtain
\begin{align}
N_\ast &= \frac{1}{M_{\rm Pl}^2}\int^{\phi_\ast}_{\phi_{\rm e}}\left( 1+\xi M_{\rm Pl}^{-2}V\right)\frac{V}{V_\phi} {\rm d}\phi = \frac{1}{M_{\rm Pl}^2}\int^{\phi_\ast}_{\phi_{\rm c}}\frac{V}{V_\phi} {\rm d}\phi + \frac{1}{M_{\rm Pl}^2}\int^{\phi_{\rm c}}_{\phi_{\rm e}}\left( 1+\lambda M_{\rm Pl}^{-2}V\right)\frac{V}{V_\phi} {\rm d}\phi, 
\end{align}
where $\phi_{\rm e}$ denotes the field value at $t=t_{\rm e}$. Since inflation ends shortly before the field reaches $V=0$, we approximate $\phi_{\rm e}$ by the field value satisfying $V=0$, such as $\phi_{\rm e}=0$ for the monomial potential and the $\alpha$-attractor E-model.

We first consider the monomial potential, for which the {\it e}-folding number becomes
\begin{align}
N_\ast = &\frac{1}{M_{\rm Pl}^2}\int^{\phi_\ast}_{\phi_{\rm c}} \frac{M^2}{p\phi}\left(1 + \frac{\phi^2}{M^2} \right)^{1-\frac{p}{2}}\left[ \left( 1 + \frac{\phi^2}{M^2} \right)^{\frac{p}{2}}- 1\right] {\rm d}\phi  \nonumber \\
& + \frac{1}{M_{\rm Pl}^2}\int^{\phi_{\rm c}}_{\phi_{\rm e}}\left\{ 1+ c\left[ \left( 1 + \frac{\phi^2}{M^2} \right)^{\frac{p}{2}}- 1\right] \right\} \frac{M^2}{p\phi}\left(1 + \frac{\phi^2}{M^2} \right)^{1-\frac{p}{2}}  \left[ \left( 1 + \frac{\phi^2}{M^2} \right)^{\frac{p}{2}}- 1\right] {\rm d}\phi \nonumber \\
\simeq & \frac{\phi_\ast^2}{2M_{\rm Pl}^2p}\left[ 1 + \frac{2}{p-2}\left( \frac{\phi_\ast}{M}\right)^{-p} \right]  + \frac{c\phi_{\rm c}^2}{M_{\rm Pl}^2p}\left[\frac{1}{p+2}\left( \frac{\phi_{\rm c}}{M}\right)^p -\frac{1}{p-2}\left( \frac{\phi_{\rm c}}{M}\right)^{-p} -1  \right],
\label{eq2}
\end{align}
where $c=\lambda M_{\rm Pl}^{-2}\Lambda^4$, a notation that will also be used in the following cases. The scalar spectral index and the tensor-to-scalar ratio at the pivot scale are given by
\begin{align}
n_s &= 1 - \left[{ p(p+2) \left(\frac{\phi_\ast}{M}\right)^{2p} + 2p(p-1) \left(\frac{\phi_\ast}{M}\right)^{p} } \right] { \left[ \left( \frac{\phi_\ast}{M} \right)^p -1 \right]^{-2} \left(\frac{\phi_\ast}{M_{\rm Pl}}\right)^{-2}},
\label{eq3}\\
r &= 8p^2\left(\frac{\phi_\ast}{M}\right)^{2p}\left[ \left( \frac{\phi_\ast}{M} \right)^p -1 \right]^{-2} \left(\frac{\phi_\ast}{M_{\rm Pl}}\right)^{-2}.
\label{eq4}
\end{align}
Semi-analytical predictions for $n_s$ and $r$ can then be obtained by numerically solving Eq. \eqref{eq2} and substituting the resulting $\phi_\ast$ into Eqs. \eqref{eq3} and \eqref{eq4}.

For the $\alpha$-attractor E-model, one finds
\begin{align}
N_\ast = &\frac{1}{2M_{\rm Pl}}\sqrt{\frac{3\alpha}{2}} \int^{\phi_\ast}_{\phi_{\rm c}} \left( e^{\sqrt{\frac{2}{3\alpha}}\frac{\phi}{M_{\rm Pl}}} -1 \right) {\rm d}\phi + \frac{1}{2M_{\rm Pl}}\sqrt{\frac{3\alpha}{2}} \int^{\phi_{\rm c}}_{\phi_{\rm e}}\left[ 1 +c \left( 1- e^{-\sqrt{\frac{2}{3\alpha}}\frac{\phi}{M_{\rm Pl}}} \right)^2 \right] \left( e^{\sqrt{\frac{2}{3\alpha}}\frac{\phi}{M_{\rm Pl}}} -1 \right)  {\rm d}\phi \nonumber \\
= & \frac{1}{2}\sqrt{\frac{3\alpha}{2}} \left.\left( \sqrt{\frac{3\alpha}{2}}e^{\sqrt{\frac{2}{3\alpha}}\frac{\phi}{M_{\rm Pl}}} - \frac{\phi}{M_{\rm Pl}} \right)\right|^{\phi_\ast}_{\phi_{\rm c}} + \left.\frac{3\alpha}{4}\left[ \frac{c}{2} e^{-2\sqrt{\frac{2}{3\alpha}}\frac{\phi}{M_{\rm Pl}}}  -3c e^{-\sqrt{\frac{2}{3\alpha}}\frac{\phi}{M_{\rm Pl}}} + (1+c)e^{\sqrt{\frac{2}{3\alpha}}\frac{\phi}{M_{\rm Pl}}} - (1+3c) \sqrt{\frac{2}{3\alpha}}\frac{\phi}{M_{\rm Pl}}\right]\right|^{\phi_{\rm c}}_0 \nonumber \\
\simeq& \frac{3\alpha}{4} \left(  e^{\sqrt{\frac{2}{3\alpha}}\frac{\phi_\ast}{M_{\rm Pl}}} + c e^{\sqrt{\frac{2}{3\alpha}}\frac{\phi_{\rm c}}{M_{\rm Pl}}} +\frac{3}{2}c-1 \right) - \frac{3c}{2} \sqrt{\frac{3\alpha}{2}}\frac{\phi_{\rm c}}{M_{\rm Pl}},
\end{align}
which yields
\begin{align}
\frac{\phi_\ast}{M_{\rm  Pl}} \simeq \sqrt{\frac{3\alpha}{2}}\ln\left[ \frac{4}{3\alpha} \left( N_\ast +  \frac{3c}{2} \sqrt{\frac{3\alpha}{2}}\frac{\phi_{\rm c}}{M_{\rm Pl}} \right) - c e^{\sqrt{\frac{2}{3\alpha}}\frac{\phi_{\rm c}}{M_{\rm Pl}}}  + 1-\frac{3}{2}c\right].
\end{align}
Substituting the resulting $\phi_\ast$ into the expressions for $n_s$ and $r$, we obtain
\begin{align}
n_s &= 1- \frac{8e^{-2\sqrt{\frac{2}{3\alpha}}\frac{\phi_\ast}{M_{\rm Pl}}}\left( 1 + e^{\sqrt{\frac{2}{3\alpha}}\frac{\phi_\ast}{M_{\rm Pl}}}\right)}{3\alpha\left( 1- e^{-\sqrt{\frac{2}{3\alpha}}\frac{\phi_\ast}{M_{\rm Pl}}}\right)^2} \simeq 1- \frac{8}{3\alpha}e^{-\sqrt{\frac{2}{3\alpha}}\frac{\phi_\ast}{M_{\rm Pl}}}\simeq 1 - \frac{8}{4\left( N_\ast +  \frac{3c}{2} \sqrt{\frac{3\alpha}{2}}\frac{\phi_{\rm c}}{M_{\rm Pl}} \right) - 3c\alpha e^{\sqrt{\frac{2}{3\alpha}}\frac{\phi_{\rm c}}{M_{\rm Pl}}}  + 3\alpha\left(1-\frac{3}{2}c\right)},\\
r &= \frac{64}{3\alpha \left(  e^{\sqrt{\frac{2}{3\alpha}}\frac{\phi_\ast}{M_{\rm Pl}}} -1 \right)^2 } \simeq \frac{64}{3\alpha\left[\frac{4}{3\alpha}\left( N_\ast +  \frac{3c}{2} \sqrt{\frac{3\alpha}{2}}\frac{\phi_{\rm c}}{M_{\rm Pl}} \right) - c e^{\sqrt{\frac{2}{3\alpha}}\frac{\phi_{\rm c}}{M_{\rm Pl}}}  + 1-\frac{3}{2}c \right]^2}.
\end{align}

\begin{figure*}[htb]
\centering
\includegraphics[width=0.9\textwidth]{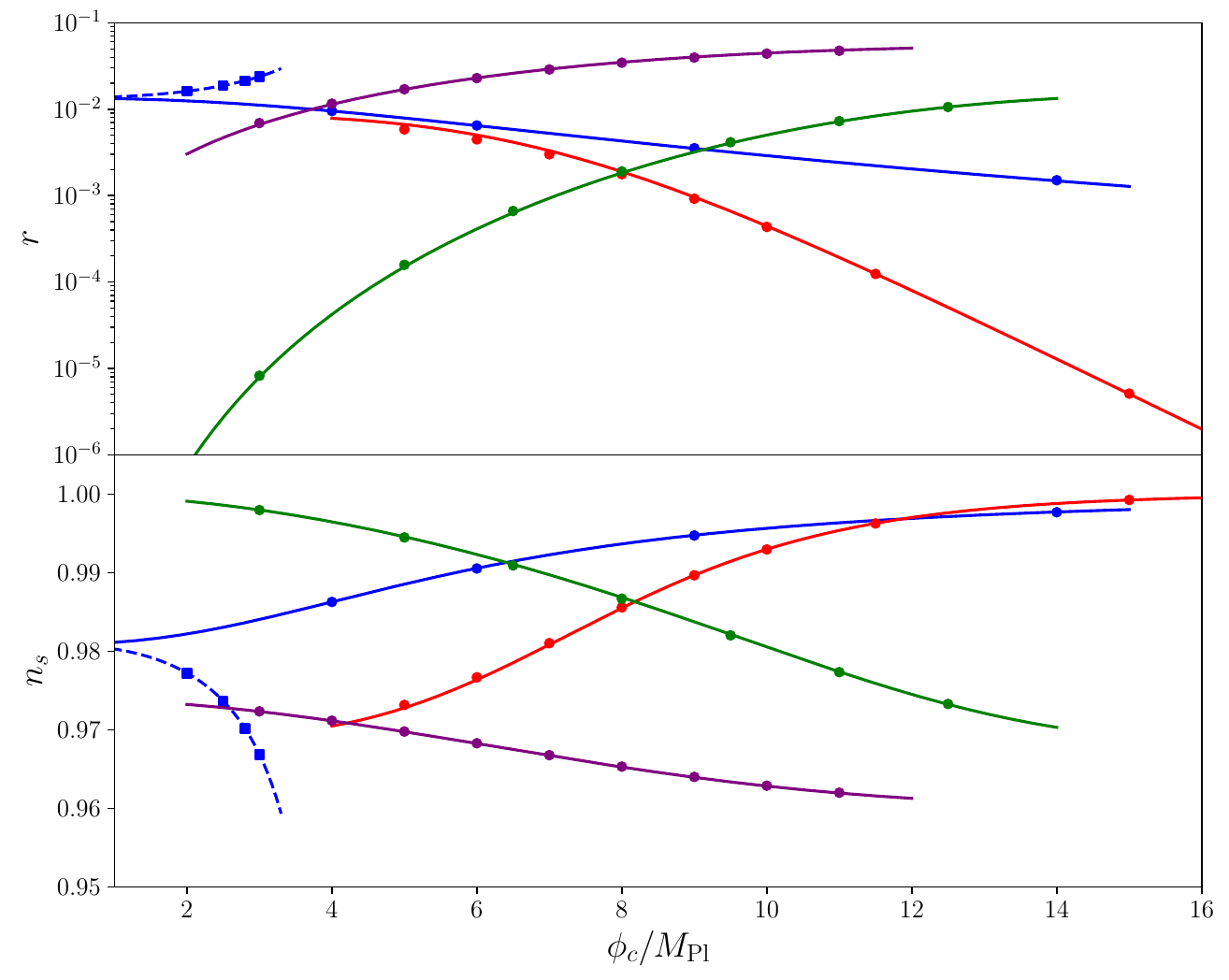}
\caption{ The (semi-)analytical predictions for $n_s$ and $r$ as functions of $\phi_c$ for monomial potential with $\{q=1/10,M=0.01M_{\rm Pl},c=-0.94\}$ (blue solid line) and $\{q=1/10,M=0.01M_{\rm Pl},c=2\}$ (blue dashed line), $\alpha$-attractor E-model with $\{\alpha=3,c=-1\}$ (red solid line), quartic hilltop potential with $\{\mu=18M_{\rm Pl},c=-1\}$ (green solid line), and natural inflation with $\{f=6.2M_{\rm Pl},c=-0.5\}$ (purple solid line). The dots and squares correspond to the numerical results shown in Fig. 3 of the main text.} \label{fig1}
\end{figure*}

We next turn to the quartic hilltop potential, for which
\begin{align}
N_\ast = &\frac{1}{M_{\rm Pl}^2} \int^{\phi_\ast}_{\phi_{\rm c}} \left(  -\frac{\mu^4}{4\phi^3} + \frac{\phi}{4} \right) {\rm d}\phi  + \frac{1}{M_{\rm Pl}^2} \int^{\phi_{\rm c}}_{\phi_{\rm e}}\left[ 1 +c \left( 1-  \frac{\phi^4}{\mu^4}\right) \right] \left( -\frac{\mu^4}{4\phi^3} + \frac{\phi}{4} \right)  {\rm d}\phi \nonumber \\
= & \left. \frac{1}{8M_{\rm Pl}^2}\left( \frac{\mu^4}{\phi^2} + \phi^2\right)\right|^{\phi_\ast}_{\phi_{\rm c}}  + \left. \frac{1}{8M_{\rm Pl}^2}\left[ (1+c) \frac{\mu^4}{\phi^2} + (1+2c)\phi^2 - \frac{c\phi^6}{3\mu^4}\right]\right|^{\phi_{\rm c}}_{\mu} \nonumber \\
=&\frac{1}{8M_{\rm Pl}^2}\left[ \frac{\mu^4}{\phi_\ast^2} + \phi_\ast^2 +  \frac{c\mu^4}{\phi_{\rm c}^2} + 2c\phi_{\rm c}^2 - \frac{c\phi_{\rm c}^6}{3\mu^4} - \left( 2+ \frac{8}{3}c\right)\mu^2 \right],
\end{align}
where we have used $\phi_{\rm e}=\mu$. Solving this equation for $\phi_\ast$, we obtain
\begin{align}
&\frac{\phi_\ast}{M_{\rm  Pl}} = \nonumber\\
&\sqrt{ 4N_\ast + \left( 1+\frac{4}{3}c \right) \frac{\mu^2}{M_{\rm Pl}^2} - \frac{c\mu^4}{2M_{\rm Pl}^2\phi_{\rm c}^2} -  \frac{c\phi_{\rm c}^2}{M_{\rm Pl}^2} +  \frac{c\phi_{\rm c}^6}{6M_{\rm Pl}^2\mu^4} - \sqrt{ -\frac{\mu^4}{M_{\rm Pl}^4} + \left[ 4N_\ast + \left( 1+\frac{4}{3}c \right) \frac{\mu^2}{M_{\rm Pl}^2} - \frac{c\mu^4}{2M_{\rm Pl}^2\phi_{\rm c}^2} -  \frac{c\phi_{\rm c}^2}{M_{\rm Pl}^2} +  \frac{c\phi_{\rm c}^6}{6M_{\rm Pl}^2\mu^4}\right]^2} }.
\end{align}
The corresponding expressions for $n_s$ and $r$ are
\begin{align}
n_s &= \frac{\mu^8-24M_{\rm Pl}^2\mu^4\phi_\ast^2-2\mu^4\phi_\ast^4-24M_{\rm Pl}^2\phi_\ast^6+\phi_\ast^8}{\left(\mu^4 - \phi_\ast^4 \right)^2},\\
r &= \frac{128M_{\rm Pl}^2\phi_\ast^6}{\left(\mu^4 - \phi_\ast^4 \right)^2}.
\end{align}

\begin{figure*}[htbp]
\centering
\includegraphics[width=1\textwidth]{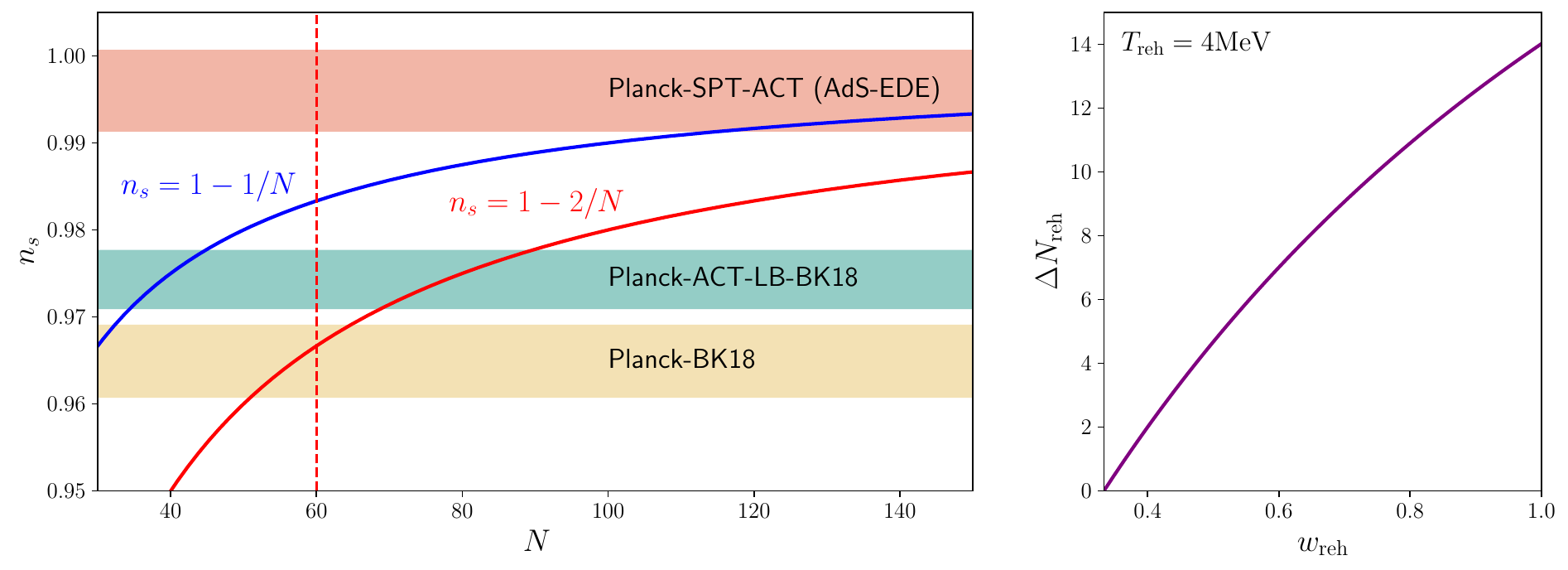}
\caption{ Left panel: The scalar spectral index $n_s$ as functions of the {\it{e}}-folding number $N$. The shaded regions show the constraints on $n_s$ at the $1\sigma$ level from Planck-BK18, Planck–ACT–LB–BK18, and Planck-SPT-ACT with AdS-EDE. Right panel: $\Delta N_{\rm reh}$ as a function of $w_{\rm reh}$ in the case of $T_{\rm reh}=4{\rm MeV}$. }  \label{fig2}
\end{figure*}

Finally, we consider natural inflation, for which
\begin{align}
N_\ast = & -\frac{f}{M_{\rm Pl}^2} \int^{\phi_\ast}_{\phi_{\rm c}} \cot\frac{\phi}{2f} {\rm d}\phi  - \frac{f}{M_{\rm Pl}^2} \int^{\phi_{\rm c}}_{\phi_{\rm e}}\left[ 1 +c \left( 1 +  \cos\frac{\phi}{f}\right) \right] \cot\frac{\phi}{2f}  {\rm d}\phi \nonumber \\
= & \left. -\frac{2f^2}{M_{\rm Pl}^2}\ln\left( \sin\frac{\phi}{2f} \right)\right|^{\phi_\ast}_{\phi_{\rm c}}  - \left. \frac{2f^2}{M_{\rm Pl}^2}\left[ (1+2c) \ln\left( \sin\frac{\phi}{2f} \right) - c\sin^2\frac{\phi}{2f}\right]\right|^{\phi_{\rm c}}_{\pi f} \nonumber \\
=& -\frac{2f^2}{M_{\rm Pl}^2}\left[ \ln\left( \sin\frac{\phi_\ast}{2f}\right) + 2c\ln\left( \sin\frac{\phi_{\rm c}}{2f}\right)\right] - \frac{2cf^2}{M_{\rm Pl}^2}\cos^2\frac{\phi_c}{2f},
\end{align}
where we have used $\phi_{\rm e}=\pi f$. Solving this equation for $\phi_\ast$, we find
\begin{align}
&\phi_\ast = 2f\arcsin\left( e^{-\frac{M_{\rm Pl}^2N_\ast}{2f^2}-c\cos^2\frac{\phi_c}{2f}}  \left(\sin\frac{\phi_{\rm c}}{2f}\right)^{-2c} \right).
\end{align}
The corresponding expressions for $n_s$ and $r$ are given by
\begin{align}
n_s &= \frac{1}{2}\left[ 1 - \frac{3M_{\rm Pl}^2}{f^2}  + \left( 1 + \frac{M_{\rm Pl}^2}{f^2} \right)\cos\frac{\phi_\ast}{f} \right]\sec^2\frac{\phi_\ast}{2f},\\
r &= \frac{8M_{\rm Pl}^2}{f^2}\tan^2\frac{\phi_\ast}{2f}.
\end{align}

Based on the above results, Fig.~\ref{fig1} shows the (semi-)analytical predictions for $n_s$ and $r$ as functions of $\phi_c$ for these four potentials. One can readily see that the (semi-)analytical results are in good agreement with the numerical ones.

We now briefly illustrate why our mechanism for shifting $n_s$ cannot be fully mimicked by post-inflationary dynamics. In the standard single-field slow-roll inflation, the scalar spectral index $n_s$ is approximately related to the {\it{e}}-folding number $N$ by
\begin{align}
    n_s \simeq 1 - \frac{\mathcal{O}(1)}{N}.
\end{align}
More specifically, one has $n_s \simeq 1 - 2/N$ for the $\alpha$-attractor E-model, and $n_s \simeq 1 - (1+p/2)/N$ for the power-law potential $\phi^p$, which reduces to $n_s \simeq 1 - 1/N$ in the limit $p\ll1$. The shift in $N_\ast$ induced by post-inflationary reheating can be written as
\begin{align}
    \Delta N_{\rm reh} = \frac{3 w_{\rm reh}-1}{4} N_{\rm reh},
\end{align}
where $w_{\rm reh}$ denotes the equation of state during reheating, and $N_{\rm reh}$ is the {\it e}-folding number during that stage. The reheating temperature is given by
\begin{align}
    T_{\rm reh} = \exp\left[ - \frac{3N_{\rm reh}(1+w_{\rm reh})}{4} \right]\left( \frac{30\rho_{\rm e}}{\pi^2g_{\rm reh}}\right)^{1/4},
\end{align}
where $g_{\rm reh}$ is the effective number of relativistic degrees of freedom at the end of reheating, and $\rho_{\rm e}$ denotes the energy density at the end of inflation. To estimate the upper bound on $\Delta N_{\rm reh}$, we take the reheating temperature to be at the lower bound imposed by BBN, namely $T_{\rm reh}=4{\rm MeV}$, and set $g_{\rm reh}=10.75$.
We further adopt $\rho_{\rm e} \simeq (10^{16}{\rm GeV})^4$, corresponding to a typical inflationary energy scale. As shown in Fig. \ref{fig2}, post-inflationary reheating with $T_{\rm reh}=4{\rm MeV}$ and $w_{\rm reh}=1$ increases $N_\ast$ by at most $\sim14$. For generic single-field slow-roll inflation, this is clearly insufficient to shift $n_s$ to the region inferred from the Planck-SPT-ACT dataset within the AdS-EDE extension of $\Lambda$CDM.

\end{document}